\documentclass{article}

\usepackage{arxiv}

\usepackage[utf8]{inputenc} % allow utf-8 input
\usepackage[T1]{fontenc}    % use 8-bit T1 fonts
\usepackage{hyperref}       % hyperlinks
\usepackage{url}            % simple URL typesetting
\usepackage{booktabs}       % professional-quality tables
\usepackage{amsfonts}       % blackboard math symbols
\usepackage{nicefrac}       % compact symbols for 1/2, etc.
\usepackage{microtype}      % microtypography
\usepackage{lipsum}		% Can be removed after putting your text content
\usepackage{graphicx}
\usepackage{natbib}
\usepackage{doi}

\usepackage{graphicx}
\usepackage{mathptmx}
\usepackage[T1]{fontenc}

\usepackage{paralist}
\usepackage{comment}

\usepackage{amsmath}
\usepackage{amsfonts}
\usepackage{amssymb}
\usepackage{amsbsy}
\usepackage{amsthm}
\usepackage{xcolor}
\usepackage{svg}
\usepackage{caption}
\usepackage{subcaption}
\usepackage[ruled]{algorithm2e}
\DeclareUnicodeCharacter{0301}{\'{e}} 
\def\E{{\mathbb E}}

\usepackage{booktabs}

\title{Structure-Function Dynamics Hybrid Modeling:\\ RNA Degradation}

\author{
\href{https://orcid.org/0000-0001-9555-7132}{\includegraphics[scale=0.06]{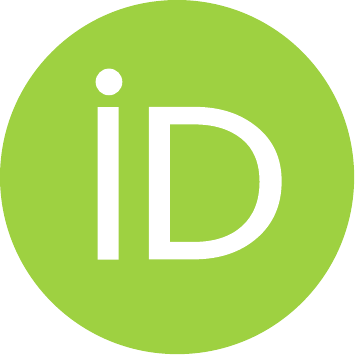}\hspace{1mm}Hua Zheng} \\
Northeastern University\\
\And
\href{https://orcid.org/0000-0001-9563-4927}{\includegraphics[scale=0.06]{orcid.pdf}\hspace{1mm}Wei Xie}\thanks{Corresponding author. Email: \url{w.xie@northeastern.edu}} \\
	Northeastern University\\
	\And
\href{https://orcid.org/0000-0001-7104-2265}{\includegraphics[scale=0.06]{orcid.pdf}\hspace{1mm}Paul Whitford} \\
	Northeastern University\\
	\And
\href{https://orcid.org/0000-0002-4214-2352}{\includegraphics[scale=0.06]{orcid.pdf}\hspace{1mm}Ailun Wang} \\
	Northeastern University\\
 	\And
        {%\includegraphics[scale=0.06]{orcid.pdf}\hspace{1mm}
        Chunsheng Fang} \\
Northeastern University\\
 	\And
  {%\includegraphics[scale=0.06]{orcid.pdf}\hspace{1mm}
  Wandi Xu} \\
	Northeastern University\\
}

\date{}

\renewcommand{\shorttitle}{\textbf{Zheng et al.}: Structure-Function Dynamics Hybrid Modeling: RNA Degradation}

\hypersetup{
pdftitle={Structure-Function Dynamics Hybrid Modeling:
RNA Degradation},
pdfauthor={Zheng et al. 2023},
}

\begin{document}
\maketitle

\begin{abstract}
RNA structure and functional dynamics play fundamental roles in controlling biological systems. Molecular dynamics simulation, which can characterize interactions at an atomistic level, can advance the understanding on new drug discovery, manufacturing, and delivery mechanisms. However, it is computationally unattainable to support the development of a digital twin for enzymatic reaction network mechanism learning, and end-to-end bioprocess design and control. Thus, we create a hybrid (``mechanistic + machine learning") model characterizing the interdependence of RNA structure and functional dynamics from atomistic to macroscopic levels. To assess the proposed modeling strategy,  in this paper, we consider RNA degradation which is a critical process in cellular biology that affects gene expression. The empirical study on RNA lifetime prediction demonstrates the promising performance of the proposed multi-scale bioprocess hybrid modeling strategy.
\end{abstract}

% keywords can be removed

\section{Introduction}
\label{sec:intro}

% =========================

% =======================

Understanding RNA structure and functional dynamics directly influences bio-drug (e.g., mRNA vaccines) discovery, manufacturing, and delivery.  %and production; for example mRNA vaccines, a promising alternative to traditional vaccines, have effectively combated various infectious diseases \citep{pardi2018mrna,buschmann2021nanomaterial}, and their potency relies on target RNA molecules' structural and functional integrity. RNA structure affects its function and RNA structure-function dynamics directly influence the manufacturing, delivery, and translation of mRNA vaccines. 
%For example, for mRNA vaccines, %a promising alternative to traditional vaccines, have effectively combated various infectious diseases. 
RNA structure affects: a) its functions and interactions
with other molecules, such as DNA, proteins, and ions; and b) the regulation of enzymatic reaction network.
For example, 
%during the mRNA delivery process,
%to individual cells, 
%the cell uptake rate relies on the binding and structural functionality of the delivery system. For the RNA translation process, structural elements, such as 3' and 5' untranslated regions (UTRs), play crucial roles in monitoring and ensuring mRNA integrity. Moreover, 
RNA structure directly affects stability, translation, %influencing storage conditions 
and delivery efficacy of RNA vaccines. The rate of RNA degradation depends on various factors, 
including pH, temperature, and ionic concentrations.

RNA structure %affects its function. 
is described in terms of three levels: primary, secondary, and tertiary. The primary structure is the nucleotide sequence of the RNA molecule, represented by four base letters (i.e., A, U, C, G). The secondary structure refers to the pattern of hydrogen bonding (base pairing) along the chain (i.e., helices), while the tertiary structure denotes the final 3D shape of the RNA molecule, determined by both secondary structure hydrogen bonding and additional nucleotide interactions. 
In recent years, there is a surge in deep learning-based algorithms for biomolecular 3D structure prediction; for example the success of AlphaFold2 \citep{jumper2021highly} has garnered the most  attention. However, the field of RNA structure prediction and structure-function dynamics modeling remains largely unexplored within the OR community, presenting potential opportunities for new insights and advancements \citep{xie2022discovery}.

The functions of RNA molecules are closely linked to their structure and dynamics.  Computer simulations, in particular molecular dynamics (MD) methods, allow structural dynamics of biomolecular systems to be investigated with unprecedented temporal and spatial resolution; see \cite{sponer2018rna}. % for a comprehensive review of the fast-developing field of MD simulations of RNA molecules). 
However, MD simulations are complex and time-consuming. 
Typically, the MD and coarse-grained simulation can only probe structure conformation change at very short time scales, i.e., $10^{-12} \sim 10^0$ second.

To efficiently model conformational dynamics and ensure scientific interpretability,
we propose a hybrid (``mechanistic + machine learning") model 
characterizing the interdependencies of RNA structure-function dynamics from atomistic  to macroscopic levels. 
This approach is based on physics-based dimensional reduction to describe some key properties: (1) interatomic interactions and potential energies quantifying global connectivity and atomic interdependencies; (2) solvation shells that approximate the aggregated effects of diffuse ions; and (3) free energy barriers and lifetimes for RNA conformational changes.
\textit{The proposed hybrid modeling strategy is general and it provides insight into energetics and dynamics of enzymatic reactions and biomolecular conformational change, which supports regulation mechanism learning and reaction rate prediction.}

To assess its performance, 
we consider RNA degradation, as measured based on unfolding times, and introduce an RNA lifetime hybrid network model (RNA-LifeTime) to quantify the structural changes of RNA during the degradation processes. By analyzing the lifetime of native contacts (defined based on interactions found in the folded molecule), RNA-LifeTime provides valuable insights into structural stability of RNA molecules under varying environmental conditions, such as temperature and ionic concentrations, and serves as a versatile probe for exploring the RNA degradation processes. 
%and accurately predicts degradation rates. 
% We employ the "Shadow" map contact algorithm as described by \cite{noel2012shadow} to identify native contacts among residues. Subsequently, we characterize the integrity of the RNA sequence by calculating the fraction of these native contacts. Then SMOG 2 is used \citep{noel2016smog,wang2022diffuse} to perform molecular dynamics (MD) simulations, and the resulting trajectories serve as the training data for RNA-LifeTime.
%By using molecular dynamics simulations and machine learning techniques, RNA-LifeTime accounts for the disruption of interatomic interactions, changes in 3D tertiary structures, and the probability of lifetime of native contacts. 
We employ the effective energy potential associated with atom-atom electrostatic interactions and solvent-mediated ionic interactions as the driving force.  
%simplified energetics for each RNA residue, along with the effective potential energy associated with electrostatic interactions and solvent-mediated ionic interactions as the driving force for RNA degradation.  
% Additionally, we introduce the concept of multi-headed Gaussians, an efficient architecture for modeling the 3D structural information of a molecule. 
%The RNA-LifeTime model offers a powerful tool for understanding RNA degradation dynamics and opens new avenues for studying biomolecular folding and function.
The proposed RNA-LifeTime can efficiently improve the prediction of RNA degradation rate.

In sum, in this paper, we made the following {contributions}: 1) We developed the RNA-LifeTime hybrid model that effectively incorporates 3D structural information and conformational dynamics of biomolecules in a diffuse ionic environment;  2)
 To the best of our knowledge, RNA-LifeTime is the first 3D molecular hybrid model capable of predicting the kinetics of RNA degradation/unfolding; 3) We develop a potential energy aggregate model employing an effective 3D spatial modeling technique called ``multi-headed Gaussians;" 4) The empirical study demonstrates the efficacy of our approach and shows that RNA-LifeTime can achieve high accuracy on RNA degradation rate estimation; %with simulation data of 14 RNA molecules; 
 and 5) By offering accurate probabilistic predictions for RNA lifetimes, our work provides valuable insights into the factors governing RNA folding processes, which establishes the groundwork for advancements in RNA-based therapeutics and diagnostics. 
 %5) The whole RNA-LifeTime framework, including code, model, and data, will be made publicly available.

The organization of the paper is as follows. In Section~\ref{sec: background}, 
we present the physics foundation for the proposed hybrid model on RNA structure-function dynamics.
%we provide the background of structure and functional dynamics of RNA and MD. 
In Section~\ref{sec:RNA degradation hybrid model}, we introduce the RNA-LifeTime model. Subsequently, we utilize MD simulations of RNA degradation/unfolding processes to evaluate the performance of our proposed approach and compare it with baseline models in Section~\ref{sec: empirical study}. We conclude this paper in Section~\ref{sec: conclusion}.

\section{RNA Structure-Function and Molecular Dynamics}\label{sec: background}

\textit{The proposed RNA structure-function dynamic hybrid model can facilitate the learning of enzymatic reaction network regulation mechanisms through MD simulations.}
We consider RNA structural dynamics (e.g., degradation/unfolding rate) as a function of environment conditions, including ion concentrations and temperature. 
At any time $t$, RNA structure evolution can be modeled with a state-action transition, where the state $\pmb{s}_t=(\mathbf{X}_t,\pmb{z}_t)$  includes the RNA structure, denoted by $\mathbf{X}_t$ (or key features characterizing RNA structure-function), and the environmental conditions denoted by $\pmb{z}_t$, %The action $\pmb{a}_t$ represents inputs that can impact the environment conditions,
\begin{equation}\label{eq.state_space}
\pmb{s}_{t+1} = f(\pmb{s}_t; \pmb{\beta}_t (\pmb{s}_t)) +\pmb{e}_t,
% \pmb{s}_{t+1} = f(\pmb{s}_t,\pmb{a}_t; \pmb{\beta}_t (\pmb{s}_t, \pmb{a}_t)) +\pmb{e}_t,
\end{equation}
 where %$\pmb{s_t}$ is the current state of the RNA, $\pmb{a_t}$ represents the current action and external influence; 
 $\pmb{\beta}_{t}(\pmb{s}_{t})$ represents the kinetic parameters (such as degradation rate) characterizing the regulation mechanism. 
 %, which depends on the current state, i.e., RNA structure and conditions. %The functional structure of $f(\cdot)$, characterizing the state dynamics, 
The residual $\pmb{e}_t$ represents the impact from other factors and model error of $f(\cdot)$. 
 
 The proposed %RNA structure dynamics 
 model is built on the scientific understanding: (1) interatomic interactions and potential energy in Section~\ref{subsubsec:RNA structure}, accounting for RNA system global connectivity and interdependencies; (2) Gaussian mixture distribution approximating the aggregate effect from hydration shells in Section~\ref{subsubsection:RNA-ion interaction}; and (3) free energy barrier crossing required for RNA conformation change in Section~\ref{subsubsection:freeEnergy-stateTransitionTime}; and (4) RNA structural dynamics balancing the driving forces induced by energy potential and thermodynamics in Section~\ref{subsec:regulationMechanism}.

\subsection{RNA Structure and Environmental Impact}% - State $\pmb{s}_t$}
\label{subsec:RNA-state}

As a cyber-physical system, an RNA system is composed of atoms with charge; 
%with short-range bonded and long-range non-bonded interactions, including hydrogen bonding, base stacking, and electrostatic interactions, van der Waals interactions; 
see Figure 1. 
The interactions between atoms from the same RNA molecule include: (1) short-range bonded interactions such as bond stretching, angle bending, and torsion, which can be related to the physical network; and (2) long-range non-bonded interactions, such as van der Waals and electrostatic forces, which can be related to the cyber network.
These interactions give rise to the potential energy that governs folding and conformational changes of the RNA structure. %So, the energy potential can reveal the preferred conformational states of a molecule, 
At the same time, atoms %constantly 
have random vibrations due to thermal energy.
\textit{Therefore, the balance of driving forces introduced by the potential energy and thermal energy influences critical pathways and reaction rates of regulatory reaction networks, e.g., RNA stability.}

\begin{figure}[ht]
%\begin{wrapfigure}{th!}{0.65\textwidth}
	\centering
	\includegraphics[width=0.85\textwidth]{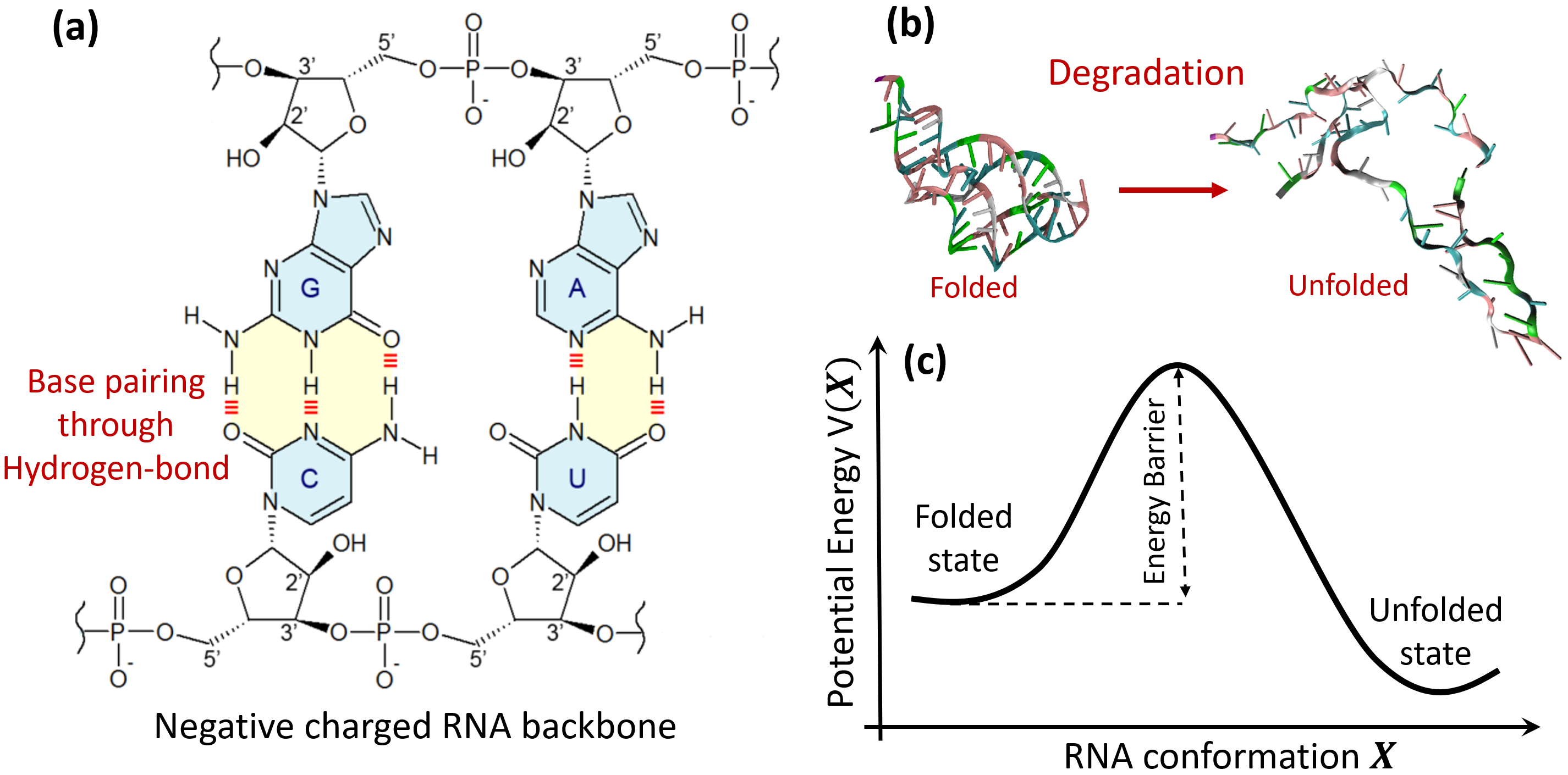}
	\caption{The RNA structural dynamics is influenced by atom interactions and thermodynamics: \textbf{(a)} Base pairing through hydrogen bond; \textbf{(b)} RNA conformational change; and \textbf{(c)} Free-energy barrier crossing for (b). %, e.g., RNA degradation.
 }
	\label{fig:RNA-Structure}
\end{figure}

\subsubsection{RNA System Structure-Function and Potential Energy}
\label{subsubsec:RNA structure}

\textit{RNA function depends on its structure}, which is typically specified in terms of primary, secondary, and the tertiary atomic structure. Primary structure defines an RNA sequence in terms of its constituent nucleotides.  The secondary structure of RNA molecule, including the 2D structure, is determined by the formation of hydrogen bonds between complementary base pairs (i.e., A=U and C=G base pairs). % also called \textit{native contact $\delta$}. 
%These canonical base pairs are particularly stable.
The 2D structure can impact the accessibility of the RNA molecule to other molecules, such as proteins and ions, and can therefore affect its function. %Typically, RNA molecules with a highly structured 2D structure may be less accessible to binding proteins, whereas RNA molecules with a more flexible 2D structure may be more accessible. 
Further building on the basic architecture defined by the 2D structure,  
the 3D structure is determined by the interactions between the atoms in the RNA molecule, and it is also important for biological functions, e.g., stability, activity, and binding specificity. 
For example, RNA molecules with a stable 3D structure may be less susceptible to degradation, while RNA molecules with a flexible 3D structure may be more dynamic and adaptable to different environments. 
 % include: the number of base pairs,CG content, free energy, and length of single-stranded segments. These variables in a), b), c), and e) impact on the driving force field which mimics the interatomic forces. 
%For 2D structure, we consider the bonding of WC canonical base pairs (A=U and C=G base pairs), which are particularly stable. For 3D structure is defined by some canonical base pair interactions and many noncanocial base pair interactions involving H-bonding. Its dynamics is conditional on 2D structure and less stable.

%\noindent\textbf{(2) Atomic Interactions and Potential Energy .} 

%\subsubsection{Atomic Interactions and Potential Energy}
%\label{subsubsec:potentialEnergy}

%Then, we describe atomic interactions and the potential energy determining RNA structure dynamics. 
Atom-atom bonded and non-bonded interactions determine the potential energy, denoted by $V$, 
%As we know, the interactions between atoms in RNA molecules can be mathematically represented by a potential energy function defined as $V$. Moreover, 
%Thus, the potential energy, denoted by $V$, can be written as,
%can be broken down into short-range interactions named $V_{short}$, and long-range interactions called $V_{long}$:
\begin{equation}
    V = V_{bonded} + V_{non-bonded}. \nonumber
\end{equation}
which induce the constraints on RNA structural dynamics and are associated with an energy barrier to undergo  a conformational change (see Figure 1).
The bonded interactions play a crucial role of defining the precise stereochemistry of the molecule. These short-range interactions are typically modeled using harmonic, or periodic potentials, as implemented using semi-empirical potentials, such as the AMBER force field, which take into account the interactions between bonded and angles and dihedrals potentials. % in the RNA molecule.

The non-bonded interactions, occurring between atoms that are distant in the RNA sequence, are typically modeled using distance-dependent potentials, such as the Lennard-Jones potential and the Coulomb potential. The Lennard-Jones potential represents the van der Waals (VDW) interactions between atoms, while the Coulomb (C) potential represents the electrostatic interactions between charged atoms: $V_{non-bonded}=V_C  + V_{VDW}$.
% \begin{equation}
%      V_{non-bonded}=V_C  + V_{VDW}.
% \end{equation}
%The full expression for the electrostatic potential of a molecular system, with the $i^{th}$ nucleus at position $R_{i}$ and having charge $Z_{i}$, is
%$
%  \varphi(r)=\sum_{i \in {nuclei}} \frac{Z_{i}}{||r-R_{i}||}-\int \frac {\Psi^2(r')}{||r'-r||} dr'.
%$
The Coulomb potential is induced by any pair of atoms $i$ and $j$ with charge $q_i$ and $q_j$,%. The charge of each atom was derived from the Amber99sb-ildn force field, accounting for the impact of chelated (i.e., inner shell) ions are partially dehydrated, which allows them to form strong direct contacts with RNA. It is defined as following: 
\begin{equation}\label{eq: Coulomb potentials}
   V_C=\sum_{ij \in \mathcal{R}} \frac {q_{i}q_{j}}{4\pi \epsilon \epsilon_{0}r_{ij}},
\end{equation}
where  $r_{ij}$ is the interatomic distance, $\mathcal{R}$ represents the set of atoms in the RNA system,
$\epsilon$ is the dielectric constant
for water, and $\epsilon_{0}$ is the permittivity of free space.

\subsubsection{RNA-Ion Interactions and Effect on Energy Potential}
\label{subsubsection:RNA-ion interaction}

\begin{figure}[ht]
%\begin{wrapfigure}{th!}{0.65\textwidth}
	\centering
	\includegraphics[width=0.8\textwidth]{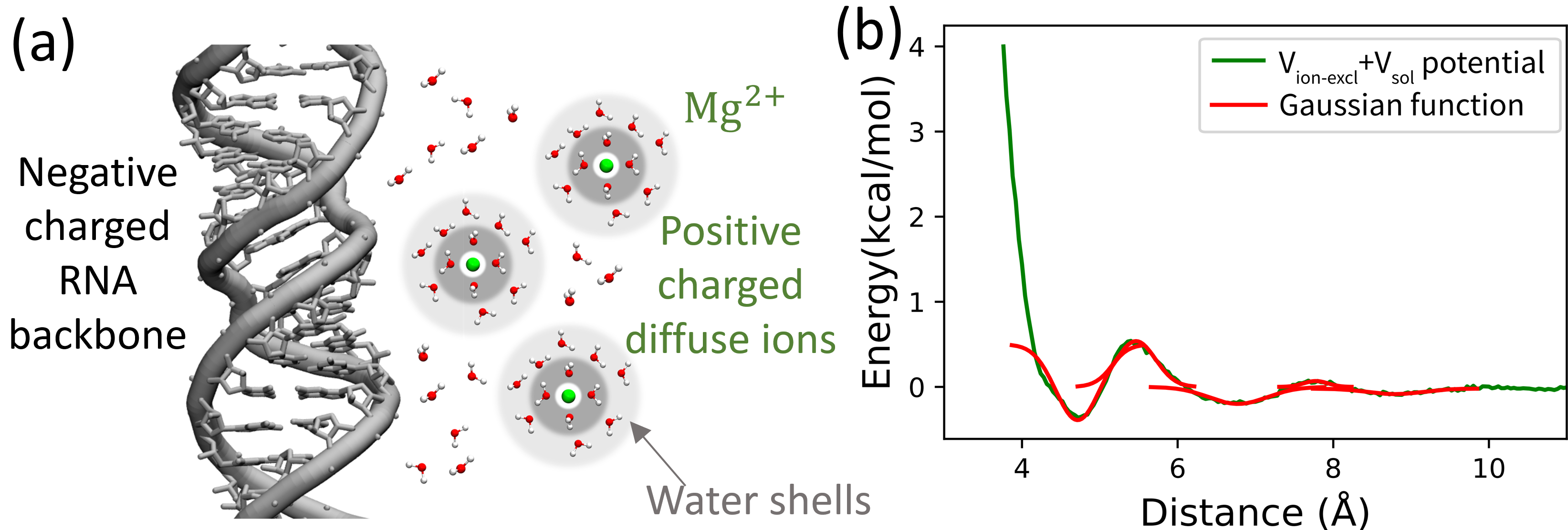}
	\caption{\textbf{(a)} An illustration of RNA and diffuse ion interactions; and \textbf{(b)} Gaussian mixture distribution (red line) approximates the energy potential of ions excluding Coulomb potential, $V_{ion-excl} + V_{sol}$ in eq.~(\ref{eq.V_E}) (green line). This plot is adapted from \cite{wang2022diffuse}.
 } \label{fig:ion-RNA}
\end{figure}

RNA structure can be impacted by the ion concentration environment through {ion-mediated electrostatic interactions}. 
Here we consider ion-RNA interactions and study the impact on the energy potential \citep{wang2022diffuse}. 
Positively charged ions (e.g., $\mbox{Mg}^{2+}$) can interact with negatively charged RNA backbone through Coulomb electrostatics (see Figure~\ref{fig:ion-RNA}).
%which can  influence the effective potentials. % and the free-energy barriers on RNA structure conformational change. 
Due to the creation of \textit{inner- and outer- hydration shells}, % induced by water molecules, 
the solvent environment around RNA is generally described
as containing chelated and diffuse ions. In the inner-shell, chelated 
ions %, denoted by the set $\mathcal{M}_i$, 
are partially dehydrated, which allows them to form
strong direct contacts with RNA. 
As a result, chelated ions can remain bound on RNA for millisecond time scale. 
In the outer-shell, diffuse ions
%, denoted by the set $\mathcal{M}_o$, 
remain fully
hydrated (e.g., $\mbox{Mg(H}_2\mbox{O})^{2+}_6$) and associate less strongly with RNA. Despite
the transient and weak influence of individual diffuse ions, the behavior of a diffuse ion is primarily
determined by longer-range electrostatic interactions, and their
collective effect on RNA structure can be significant. 
Therefore, we consider the \textit{aggregated effects} of ion-RNA interactions on energy potential due to the existence of inner- and outer-shells during the RNA structural dynamics hybrid modeling development.  

To account for ionic interaction effect, the electrostatic energy potential ($V_E$) includes updated %direct 
Coulomb interactions ($V_C^\prime$), %, effective excluded volume potentials for diffuse ions ($V_{ion−excl}) and 
and effective potentials that describe ionic solvation
effects ($V_{sol}$) and excluded volume of ions ($V_{ion-excl}$); see \cite{wang2022diffuse} for details,
\begin{equation}
    V_E = V_C^\prime+ \left[ V_{sol}+V_{ion-excl} \right] 
    = 
    \sum_{ij\in \mathcal{R} 
    %\cup \mathcal{M}_i\}
    } 
    \frac {q_{i}^\prime q_{j}^\prime}{4\pi \epsilon \epsilon_{0}r_{ij}} 
    % +\sum_{ij\in \mathcal{R}} \frac{A}{r_{ij}^{12}}
    + \left[\sum_{ij \in \mathcal{R} 
    %\cup \mathcal{M}_o\}
    } 
    \left(
    \sum_{k=1}^K 
    B^{(k)}e^{-C^{(k)} \left[ r_{ij}-R^{(k)}\right]^2}
    \right)\right].
    \label{eq.V_E}
\end{equation}
The energy potential $V_C^\prime$ represents the direct Coulomb interactions with updated charges $q_i^\prime$ and $q_j^\prime$ accounting for the impact from chelated %diffuse 
ions with strong contacts with RNA molecule. For simplification, the mean-field approach is used here. The potential energies related to solvent-mediated ionic interactions and excluded volume of ions, represented by $V_G \equiv V_{sol}+V_{ion-excl}$, are modeled by a sum of Gaussians, characterizing the aggregated effect of diffuse ions in the hydration shells.
It accounts for up to three outer hydration shells, as shown in Figure~\ref{fig:ion-RNA} (adapted from \cite{wang2022diffuse}).

For RNA stability analysis, the energy function used to study the difference of energy between folded and unfolded states can be simplified. 
%This energy function can then be applied to the RNA structure in folded and unfolded states. When the calculation is complete, the two are subtracted to estimate the driving force. But, in the realistic, there are some situations we need to consider. 
First, we can ignore the change in bond potential because the bonds are too strong to break unless the temperature is extremely high. %because the parameters of harmonic bond potential in SMOG makes the bond are too strong to break unless the temperature is extremely high. 
Second, the harmonic angle potential is similar and it is less related to the native contact definition used to measure RNA folding. Third, the Coulomb potential is a key driving force of inter-atomic interaction that contribute to native contacts. 
% At last, to simply our model, we do not consider the Van Der Waals potential. 
\textit{Therefore, the energy barrier used in the paper for RNA degradation rate estimation is related to $V_{C}^\prime$ and $V_{sol}+V_{ion-excl}$.}
% ; as shown in eq.~(\ref{eq.V_E}).}

\subsubsection{Free-Energy Barrier Crossing for Conformational Change}
\label{subsubsection:freeEnergy-stateTransitionTime}

The potential energy $V$ imposes constraints on the RNA structure by favoring conformations that have lower local potential energy. 
When we have the conformational change from state $\mathbf{X}_1$ to $\mathbf{X}_2$, the \textit{energy barrier}, defined as $\Delta G = \max V(\mathbf{X})-V(\mathbf{X}_1)$, represents the difference in the energy between the starting state $\mathbf{X}_1$ and the maximum potential energy, denoted by $\max V(\mathbf{X})$, occurring during the transition process; as shown in Figure~\ref{fig:RNA-Structure}.
% $\Delta G=V(X_1) - V(X_2)$ , say from state $X_1$ to $X_2$, becomes $\Delta G=V(X_1) - V(X_2)$.  
% ∆GTSE is the difference in the free energy of the A/T ensemble and the transition state ensemble (TSE) and Ca is the barrier-crossing attempt frequency
Then, the rate of accommodation denoted by $k_a$ (similar to the rate acceleration in enzymatic
reactions) and the mean-first passage time (e.g., the lifetime of contacts transitioning from the folded to the unfolded state) denoted by $T$ have the following relationship with $\Delta G$,
\begin{equation}\label{eq: relationship between lifetime and energy}
    k_a = \frac{1}{T} %\approx 
    \propto \frac{1}{C_a} \exp\left(\frac{\Delta G}{k_B\mathbb{T}} \right),
\end{equation}
where $k_B$ is the Boltzmann constant, $\mathbb{T}$ is the temperature, and $C_a$ is the
barrier-crossing attempt frequency. Methods for estimating these prefactors or coefficients are described in \cite{whitford2010connecting}.
For the RNA degradation process with monotonic change in free energy, we have $\Delta G =  V(\mathbf{X}_1)-V(\mathbf{X}_2)$.
\textit{The energy barrier, referring to the amount of energy that must be supplied to a biomolecular system in order for it to undergo a particular transformation, is influenced by factors, such as temperature and ion concentration.}

The proposed hybrid model can be extended to estimate reaction rates, learn regulation mechanisms, and support optimal learning/control for enzymatic reaction networks, where free-energy barrier height depends on the capability of enzymes and environmental conditions. 
%by using molecular dynamic simulation, even though this paper focuses on RNA degradation rate estimation.
Basically, for general enzymatic reactions, the most important contribution to catalysis comes from the reduction of the 
%activation free energy or 
free-energy barrier by electrostatic effects to simulate the rate acceleration in enzymatic
reactions \citep{Jordi_2001}.
%Enzyme can reduce the free-energy barrier to accelerate the chemical reaction rate. 
The catalytic power of enzymes depends on factors, such as temperature, pH, and ion concentration.

%At last we talk about the Final conformation. 
%Once the RNA molecule has passed through the transition state, it can reach the final conformation  characterized by a stable arrangement of base pairs and RNA backbone. This allows the RNA molecule to perform its biological function, such as catalyzing chemical reactions or regulating gene expression.
%The proposed hybrid model could provide insight into the energetics and dynamics of enzymatic reactions. 
%these reactions by calculating the potential energy changes that occur during the reaction. This can advance the understanding of how enzymes catalyze reactions at the molecular level. 

% ===================
\subsection{RNA Structure-Function Dynamics and Thermodynamics}
\label{subsec:regulationMechanism}

Langevin dynamics is used to determine RNA structure evolution accounting for: a) regulation from the potential energy gradient; and b) thermal energy modelled by Brownian motion due to \textit{temperature effects}. For any RNA system composed of $N^\prime$ atoms with masses $\mathbf{m}$ and coordinates $\mathbf{X}_t$,
Langevin equation states, 
\begin{equation}
\label{eq.LangevinDynamics}
    \mathbf{m}\ddot{\mathbf{X}}_t = -\nabla V(\mathbf{X}_t) - \gamma\dot{\mathbf{X}}_t+\sqrt{2\gamma k_B \mathbb{T}} \mathcal{G}_t
\end{equation}
where the gradient $-\nabla V(\mathbf{X}_t)$  over the energy potential gives the driving force calculated from the atoms interaction potentials, $\dot{\mathbf{X}}_t$ is the velocity, $\ddot{\mathbf{X}}_t$ is the acceleration, $\gamma$ is friction coefficient, $\mathbb{T}$ is the temperature, $k_{B}$ is Boltzmann's constant, and $\mathcal{G}_t$ is a delta-correlated stationary Gaussian process with zero mean representing thermal fluctuations.
RNA structural dynamics can be studied using a variety of computational techniques, including MD simulations, Monte Carlo simulations, and coarse-grained models.

\section{RNA Degradation Dynamics Hybrid Modeling}
\label{sec:RNA degradation hybrid model}

%The backbone of an RNA sequence is formed by a chain of ribose sugar molecules linked together by phosphodiester bonds. It provides the structural framework for RNA molecules and determines their overall shape. 

While the RNA backbone typically remains stable during the degradation process, the disruption of interatomic interactions can lead to changes in the secondary and tertiary structures of RNA. \textit{To quantify these structural changes, native contacts of residues or atoms are used in this study.} In RNA folding and degradation processes, they refer to the natural interactions, such as base-pairing and ionic interactions, between the residues or atoms of RNA molecules in their folded 3D structure.
Native contacts are defined by using the Shadow Contact Map algorithm \citep{noel2012shadow} with cutoff parameters from \cite{wang2022diffuse}. 
% To capture the dynamics of biomolecular folding and function, we utilized the "Shadow" map contact definition. The contact map is a binary symmetric matrix that indicates which atom pairs are given attractive interactions, please refer to \citep{noel2012shadow} for a more detailed definition. 
In addition, the fraction of native contacts is used to measure the deviation from the native folded state of RNA structure through MD simulations \citep{wang2019limits}. By identifying changes in the native contacts, our model provides insights into how the structural integrity of RNA molecules changes during the degradation process.

\noindent\textbf{Notation:} The numbers of simulation trajectories, types of native contact, and environmental features are denoted by $N_s$, $C_t$, and $C_z$ respectively. We denote the number of residues in the input primary sequence by $N$. Sequences of varying lengths were padded with zeros at the end to ensure they were of equal length. We use $\odot$ for the element-wise multiplication, $\otimes$
for the outer product, and $\oslash$ for element-wise division. We use $[x]$ to denote the sequence of positive integers from $1$ to $x$, where $x$ is an integer and $\lceil x \rceil$ to denote the smallest
integer greater than or equal to $x$.
We denote the standard Dropout \citep{srivastava2014dropout} with the operator $\mathtt{Dropout}_x$, where $x$ is the dropout rate, i.e., the probability setting an entry to zero. We use $\mathtt{L}(\cdot)$ for a linear transformation with
a weight vector or matrix, denoted by $\pmb{w}$ and $W$, and a bias vector $\mathbf{b}$. The feedforward network is two-layer fully connected with ReLu activation function, $$\mathtt{FeedForward}(\mathbf{X};\pmb{b}_3, \pmb{b}_4,\mathbf{W}_3,\mathbf{W}_4)=\pmb{b}_4 +\mathbf{W}_4^\top\mathtt{ReLu}(\pmb{b}_3 + \mathbf{W}_3^\top \mathbf{X}) \text{ with ReLu(x)=$\max(0,x)$}.$$
We use $\mathtt{BatchNorm}$ for the batch normalization \citep{ioffe2015batch} such that the mean and standard deviation are calculated per dimension over the mini-batches.

% ================================
\subsection{Native Contact and Its Lifetime}

\begin{figure}[bht!]
	\centering
\includegraphics[width=0.8\textwidth]{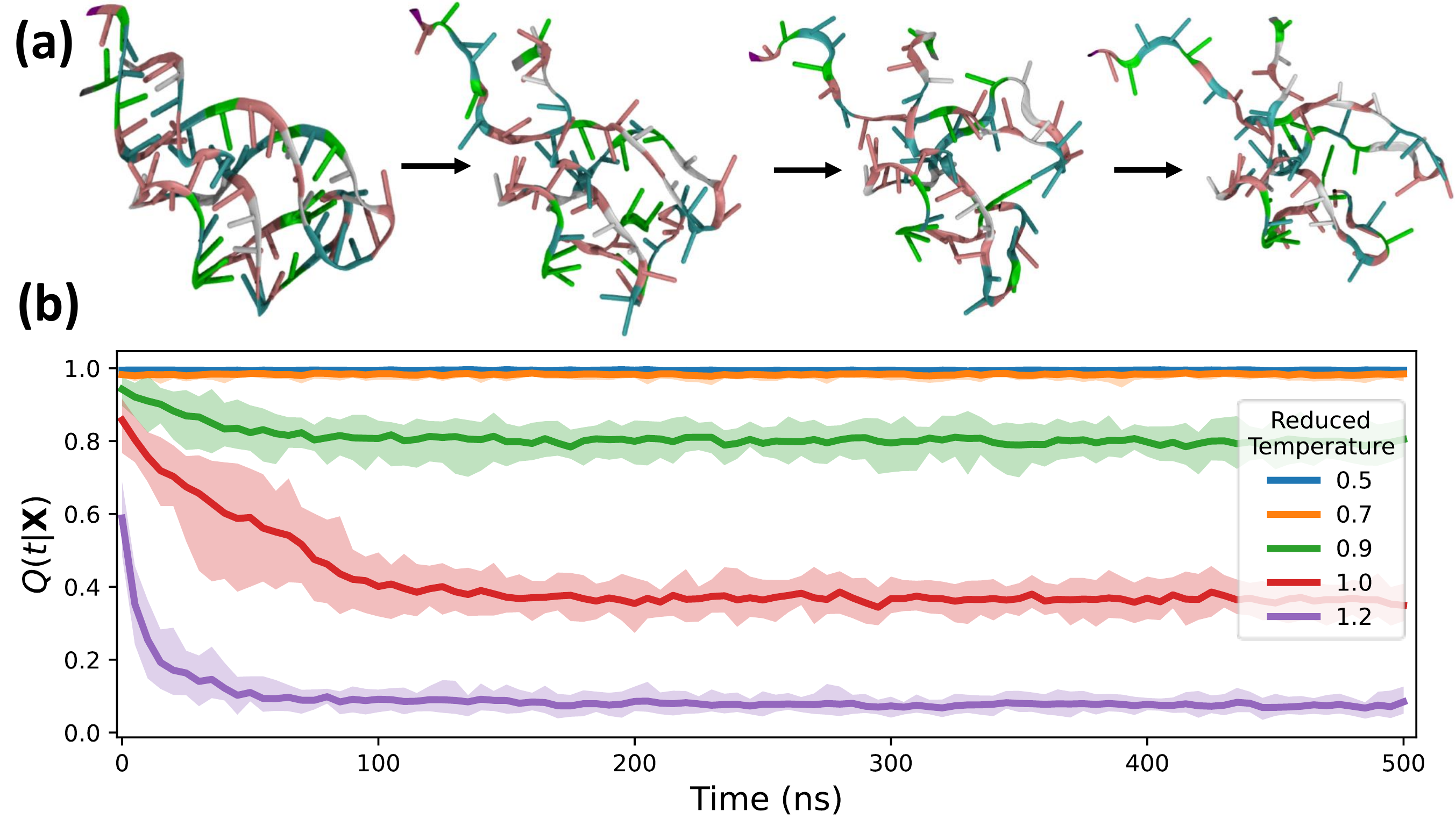}
	\caption{\textbf{(a)} The degradation of A-riboswitch-adenine complex RNA (PDB 1Y26) in the first 20 ns of MD simulation. \textbf{(b)} The fraction of native contacts $Q(t|\mathbf{X}_0,\pmb{z})$ is changing with time at varying reduced temperatures. The reduced temperature is a unit used in MD simulation where 0.5 corresponds to room temperature, and 1.0 corresponds to 350K.
 % The sequence stays in the folded state when the temperature is below 1.0 and starts around $\mathbb{T}=1.0$, suggesting the presence of a free energy barrier.
 } 
	\label{fig: degradation process}
\end{figure}

The lifetime of a native contact has been introduced by \cite{best2013native} to quantify the importance of an individual inter-residue contact in the protein folding mechanism. We adopt this concept and apply it to RNA degradation process (Figure~\ref{fig: degradation process}) by studying the trajectory of fraction of native contacts, i.e., the ratio of the number of contacts present in the current structure to the number of contacts in the native structure.

%Native contacts have varying lifetimes, which affects the fraction of native contacts during the degradation process. 

As the RNA molecule degrades, the fraction of native contacts reflects the percentage of the sequence that remains integrated. 
Let $\mathbf{X}_0$ represent the initial native conformation of an RNA sequence of length $N$, with each element encoding location, base, and environmental condition of the residue. For simplification, suppose the environmental conditions (i.e., temperature, ionic concentration) are fixed, denoted by $\pmb{z}$.
%Let the vector $\pmb{z}$ represent the environmental conditions such as temperature and ionic concentration.
Let $T_{ij}$ represent the lifetime of native contact pair $(i,j)$ and $r_{ij}$ the distance between residue $i$ and $j$ for any $i,j\in [N]$. %for a given sequence conformation $\mathbf{X}$, 
The probability of the native contact pair $(i,j)$ remaining integrated until time $t$ is modelled by
%https://www.overleaf.com/project/641230f2bba35a7fd0b01fbcd 
\begin{equation}\label{eq: probability of lifetime of native contacts}
    \mbox{Pr}(T_{ij}>t|\mathbf{X}_0,\pmb{z})\approx\frac{1}{1+e^{\beta (r_{ij}(t|\mathbf{X}_0,\pmb{z})-\lambda r_{ij}(0|\mathbf{X}_0,\pmb{z}))}} \ \ \text{ (\textbf{Lifetime Probability})},
\end{equation}
where $\beta=50$ \AA\textsuperscript{-1} is a smoothing parameter %taken to be  
and the factor $\lambda=1.2$ accounts for the thermodynamic fluctuation when contact is formed 
\citep{best2013native}. Let $\mathbb{C}(\mathbf{X}_0)$ represent the set of native contact pairs for initial conformation $\mathbf{X}_0$. Then, following \cite{best2013native}, we define the integrity function of the RNA molecule at any time $t$ as
\begin{equation}\label{eq: fraction of native contacts}
    Q(t|\mathbf{X}_0,\pmb{z})\equiv
    %\defeq 
    \frac{1}{|\mathbb{C}(\mathbf{X}_0)|}\sum_{(i,j)\in\mathbb{C}(\mathbf{X}_0)} \mbox{Pr}(T_{ij}>t|\mathbf{X}_0,\pmb{z}) \ \ \text{(\textbf{Integrity Function})}.
\end{equation}
%Notice that eq.~\eqref{eq: fraction of native contacts} is the same as the fraction of native contacts defined in \cite{best2013native}. 
Thus, we interpret $Q(t|\mathbf{X}_0,\pmb{z})$ 
%in two different ways: (1) 
as the averaged probability of native contacts remaining by time $t$.
%, and (2) the ratio of the number of contacts that are present in the native structure to the total number of possible contacts in the RNA molecule. 
As the key characteristics of RNA integrity, the probability \eqref{eq: probability of lifetime of native contacts} will be used as the \textbf{output} of our predictive model.

% \begin{remark} A method for estimating the lifetime probability, as shown in equation \eqref{eq: probability of lifetime of native contacts}, involves modeling the distance $r_{ij}(t|\mathbf{X})$ between any pair of contacts $(i, j) \in \mathbb{C}$ at varying time steps. This technique, known as the Markov state model, has been extensively investigated \citep{wu2018deep,mardt2018vampnets,noe2020machine}. However, these models may struggle to accurately predict the atom/residue positions during degradation processes. These interactions abruptly destabilize when the temperature reaches a critical point. As native contacts break, atoms gain freedom of movement, leading to unpredictable atomic locations and varying distances between pairs.
% \end{remark}

% the ratio of the number of contacts that are present in the native structure to the total number of possible contacts in the RNA molecule. A contact is considered to be present if the distance between two atoms in the predicted structure is within a specified cutoff distance from the corresponding atoms in the native structure.
% Contacts are defined using the Shadow Contact Map algorithm67 with a
% 6 Å cutoff and a 1 Å shadowing radius.

\subsection{RNA Lifetime Modeling}\label{subsec: lifetime model}
% Consider an length-$N$ RNA sequence $\pmb{s}=(s_1,s_2,\ldots,s_N)$ with $N_c$ number of unique native contacts. The type of native contact between residue $i$ and $j$ is one-hot encoded by a binary vector $\pmb{c}_{ij}$ with each element indicating the presence or absence of the corresponding native contact type. The total number of native contact types is $K=|\pmb{c}_{ij}|$. 

% Consider $N_{traj}$ independent RNA trajectories $\left\{(\mathbf{X}_n,\mathbf{Y}_n)\right\}_{n=0}^{N_{traj}}$, each with covariates $\mathbf{X}_n = (\mathbf{D}_n, \mathbf{C}_i, \pmb{z}_i)$ and a sequence of targets $\mathbf{Y}_i(t)$ measured at timestep $t=0,1,\ldots, T$. Specifically, $\mathbf{Y}_i(t)$ represents the contact integrity matrix at timestep $t$ 

In this study, the native conformation of the RNA molecule is represented by the inter-residue distances and their contact types, i.e., $\mathbf{X}_0=(\mathbf{D}, \mathbf{C})$, where the $(i,j)$-entry of the matrix $\mathbf{D}$ represents the pair distance between residues $i$ and $j$ in the folded native state; the $(i,j, k)$-entry of the tensor $\mathbf{C}$ represents the $k$th type of contact between residues $i$ and $j$ in the folded native state. We consider \textit{input} as the initial conformation of a RNA molecule and the environmental conditions, that is $\mathbf{S}_0 = (\mathbf{X}_0, \pmb{z})$.
%a sequence of targets $\mathbf{Y}(t_\ell)$ measured at time $0\leq t_\ell\leq H$ for $\ell\in[L]$. 
The input $\mathbf{S}_0$ corresponds to the initial state of the RNA system in eq.~\eqref{eq.state_space}, i.e., $\pmb{s}_0=\mbox{vec}({\mathbf{S}_0})\equiv\left(\mbox{vec}(\mathbf{D}), \mbox{vec}(\mathbf{C}), \mbox{vec}(\pmb{z})\right)^\top$. 
For simplification, suppose the environmental conditions $\pmb{z}_t$ are fixed during the RNA degradation process; thus drop $t$ subscript.
At any time $t$, the \textit{output} of interest is the lifetime probability matrix, \begin{equation*}
    \mathbf{Y}(t)=[y_{ij}(t)]_{N\times N} ~~~ 
    \text{ where } y_{ij}(t)=\begin{cases}
\mbox{Pr}(T_{ij}>t|\mathbf{S}_0) & \text{if } (i,j)\in \mathbb{C}(\mathbf{X}_0); \\
0 & \text{otherwise.}
\end{cases}
\end{equation*} 
Then, we can calculate the observed fraction of native contacts by $Q(t|\mathbf{S}_0)=\frac{1}{|\mathbb{C}(\mathbf{X}_0)|}\sum_{(i,j)\in\mathbb{C}(\mathbf{X}_0)} y_{ij}(t)$. 

% The connection between the lifetime of a native contact and energy difference has been established by \cite{bryngelson1989intermediates,whitford2010connecting} in the form of
% \begin{equation}\label{eq: relationship between lifetime and energy}
%     \log R_0 T \propto \frac{\Delta G}{k_B \mathbb{T}}
% \end{equation}
% where $R_0$ is some position-independent reference diffusion constant, $\Delta G$ is the difference in the free energy of folded and unfolded states, $k_B$ is the Boltzmann constant and $\mathbb{T}$ is the temperature. 

Through connecting the lifetime of a native contact with the potential energy difference in eq.~\eqref{eq: relationship between lifetime and energy}, we can express the logarithm of the lifetime $T_{ij}$ of a native contact $(i,j)$ as a function of potential energy
barrier for the RNA degradation, represented by $\Delta G_{ij}$, i.e.,
\begin{equation*}
    \mathtt{L}(\log T_{ij}) = \frac{\Delta G_{ij}(\mathbf{S}_0)}{\mathtt{L}_{\mathbb{T}}(\mathbb{T})}+\sigma_{ij}(\pmb{z}) W,
\end{equation*}
where $\sigma(\pmb{z})>0$ is a scale parameter function, $\mathtt{L}(\cdot)$ and $\mathtt{L}_{\mathbb{T}}(\cdot)$ are linear functions of the logarithm of time $\log T_{ij}$ and temperature $\mathbb{T}$, and the noise $W$ follows a standard normal distribution. The term $\sigma_{ij}(\pmb{z}) W$ is introduced to account for %the approximation error induced by eq.~(\ref{eq: relationship between lifetime and energy}) and 
the random motion induced by thermodynamics; see eq.~(\ref{eq.LangevinDynamics}).

Then the lifetime probability of the native contact $(i,j)\in\mathbb{C}(\mathbf{X}_0)$ is modeled by
\begin{align*} 
    H_{ij}(t|\mathbf{S}_0) & \equiv \mbox{Pr}({T}_{ij} \geq t|\mathbf{S}_0) =\mbox{Pr}\left(\mathtt{L}(\log T_{ij})\geq \mathtt{L}(\log t )|\mathbf{S}_0\right)= \mbox{Pr}\left(W \geq \frac{\mathtt{L}(\log t)-\Delta G_{ij}(\mathbf{S}_0)/\mathtt{L}_{\mathbb{T}}(\mathbb{T})}{\sigma_{ij}(\pmb{z})}\right) \nonumber \\ 
   &=1-\Phi\left(\frac{\mathtt{L}(\log t)-\Delta G_{ij}(\mathbf{S}_0)/\mathtt{L}_{\mathbb{T}}(\mathbb{T})}{\sigma_{ij}(\pmb{z})}\right), \text{ if $W\sim \mathcal{N}(0,1)$}.
   \label{eq: lifetime  probability model for one contact}
 \end{align*}
 % In practice, we reformulate this lifetime probability \eqref{eq: lifetime  probability model for one contact} by relaxing the output as a linear function of the logarithm of time, that is, $ H_{ij}(t|\mathbf{S}_0)=1-\Phi\left(\frac{\mathtt{L}(\log t)-\mathtt{L}\left(\Delta G_{ij}(\mathbf{S}_0) /\mathtt{L}_{\mathbb{T}}(\mathbb{T})\right)}{b\sigma_{ij}(\pmb{z})}\right)$. Redefine $\Delta G_{ij}(\mathbf{S}_0)\defeq \mathtt{L}(\Delta G_{ij}(\mathbf{S}_0))$ and $\sigma_{ij}(\pmb{z})\defeq b \sigma_{ij}(\pmb{z})$ and 
 Assume $ H_{ij}(t|\mathbf{S}_0)=0$ if $ (i,j) \notin \mathbb{C}(\mathbf{X}_0)$. Then we can have the lifetime probability model in a matrix form 
\begin{equation}
\label{eq.H_fun}
    \mathbf{H}(t|\mathbf{S}_0) = \left[1-\Phi\left(\frac{\mathtt{L}(\log t)-\Delta\mathbf{G}(\mathbf{S}_0)/\mathtt{L}_{\mathbb{T}}(\mathbb{T})}{\sigma(\pmb{z})}\right) \right] \odot \mathbf{M}
\end{equation}
where $\Delta\mathbf{G}(\mathbf{S}_0)$ has $(i,j)$-entry $\Delta{G}_{ij}(\mathbf{S}_0)$ and $\mathbf{M}$ is the matrix where $(i,j)$-entry is one if $(i,j)\in\mathbb{C}(\mathbf{X}_0)$ and 0 otherwise. In other words, $\mathbf{M}$ is a binary matrix that masks out the elements of $\mathbb{C}(\mathbf{X}_0)$ from the original matrix. 

We conclude this section by connecting the lifetime probability model with the famous accelerated failure time (AFT) model in survival analysis \citep{lee2003statistical}. While both models assume that the covariate affects the log-time scale in an additive form, our proposed model offers a more flexible functional expression by incorporating the effects of environmental conditions and RNA structures.
% on the mean function and scale parameter.

% Both models assume that the covariate influences the log-time scale in an additive form. However, our proposed model takes a more general functional expression by allowing the mean function and the scale parameter to be a function of the environmental conditions and structures of RNA molecules. 

% This relaxation  enhances the versatility of our model and its applicability in MD.

\subsection{RNA Lifetime Hybrid Network Model}\label{subsec: model}

\begin{figure}[hbt!] 
	\centering
\includegraphics[width=1\textwidth]{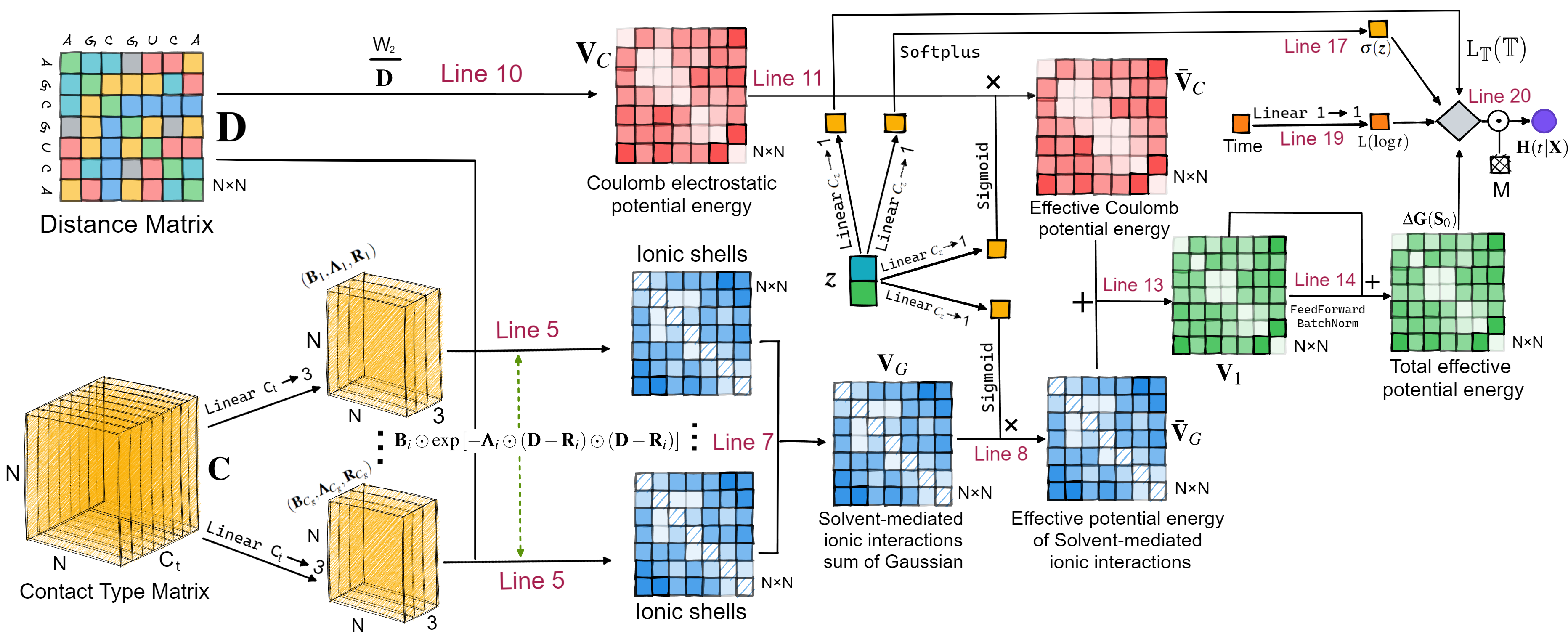}
	\caption{Model architecture. Arrows show the information flow among the various components described in this paper. Tensor shapes are shown with $N$ representing the number of residues, $C_t$ representing the number of types of native contacts, and $C_g$ representing the number of Gaussians.} 
	\label{fig: network}
\end{figure}

Algorithm~\ref{algo: model} presents the procedure on RNA structure-function dynamics hybrid modeling and RNA lifetime prediction; see 
Figure~\ref{fig: network} for the flowchart illustration.
The network model for the mean function of potential difference, denoted as $\Delta \mathbf{G}(\mathbf{S}_0)$, incorporates the potential energy associated with electrostatic interactions. 
% The functional expressions for both Coulomb potential energy and solvent-mediated ionic interactions have been adopted from \cite{wang2022diffuse}, with the added flexibility of trainable weights. The solvent-mediated ionic interactions manifest in the form of ionic shells and thus are modeled by a sum of Guassians. For each type of interaction considered, up to five Gaussians were included to describe ionic shells and unique Gaussian parameters were assigned to each type of modeled interaction.
The functional expressions for both Coulomb potential energy and the potential energy related to solvent-mediated ionic interactions and excluded volume of ions 
% $V_C^\prime+ \left[ V_{sol}+V_{ion-excl} \right]$ in eq.~(\ref{eq.V_E}),
are derived from \cite{wang2022diffuse}, with the trainable weights.

The potential energy related to solvent-mediated ionic interactions and excluded volume of ions is represented by a sum of Gaussians (Algorithm~\ref{algo: model} Line \ref{ln: for loop}-\ref{ln: sum of gaussians}). For each type of interactions considered, a Gaussian mixture with $C_g$ components is used to approximate the aggregate effect from hydration shells, with distinct Gaussian parameters designated for each specific interaction type. 
\textit{The newly introduced mechanism, referred to as ``multi-headed Gaussians'', is an efficient architecture for modeling the molecular 3D structural information.}

% The phosphate group in a nucleotide carries a negative charge of -1, while the sugar and nitrogenous base are uncharged. The overall charge of a nucleotide is therefore typically -1. 

For a coarse-grained model, the Coulomb interaction of two residues $(i,j)$ is dominated by the negatively charged phosphate group located along RNA backbone,
%(while the sugar and nitrogenous base are uncharged), 
i.e., $q_i=q_j=-1$. Therefore, the Coulomb interactions in eq.~\eqref{eq: Coulomb potentials} between any two RNA residues can be approximated as $V_{C,ij} = \frac{W_{2,ij}}{r_{ij}}$
where the trainable parameter $W_{2,ij}$, representing the term $\frac{q_{i}q_{j}}{4\pi \epsilon \epsilon_{0}}$, is a pairwise potential parameter that captures the strength of the electrostatic interaction between residues $i$ and $j$  (Algorithm~\ref{algo: model} Line \ref{ln: Coulomb electrostatic potential}). 
% In our experiments, we discovered that utilizing a more generalized form, represented as $\mathbf{V}^C =\mbox{matricize}\left(\mathtt{LinearNoBias}\left(1/\mbox{vec}(\mathbf{D}); W_2\right)\right)$, can lead to a 30\% improvement in validation loss. In this representation, $\mbox{vec}$ and $\mbox{matricize}$ denote the vectorization and matricization operators, respectively. The vectorization operator transforms a matrix into a vector, while the matricization operator converts a vector into a matrix (Algorithm~\ref{algo: model} Line \ref{ln: for loop}-\ref{ln: Coulomb electrostatic potential}).

\begin{algorithm}[th!]
\SetAlgoLined
\LinesNumbered
\textbf{Input}: The pair distance matrix $\mathbf{D}\in\mathbb{R}^{N\times N}$, native contact tensor $\mathbf{C}\in\mathbb{R}^{N\times N\times C_t}$, mask matrix $\mathbf{M}\in \{0,1\}^{N\times N}$, environmental feature $\pmb{z}\in\mathbb{R}^{C_z}$, number of contact types $C_t=10$, number of Gaussians $C_g$, and hidden layer dimension multiplier $h=4$.
\textbf{Output}: $\mathbf{H}(t|\mathbf{S}_0; \pmb\theta)\in [0,1]^{N\times N}$\\
\textbf{Function} LifetimeModel($t$, $\mathbf{D}$, $\mathbf{C}$, $\mathbf{M}$, $\pmb{z}$, $C_t=10$, $C_z=2$, $C_g=5$, $h=4$): \hfill Trainable Parameters\\
\nl \quad \ \ \textcolor{olive}{\# Solvent-mediated ionic interactions as a sum of Gaussians}\\ 
\nl \quad \ \  $\mathbf{J}=\mathbf{C} \mathbf{W}_1$ \hfill $\mathbf{W}_1\in \mathbb{R}^{C_t\times C_g\times 3}$\\
\lnl{ln: for loop}\quad  \ \ \textbf{for all} $i \in \{1,2,\ldots,C_g\}$ \textbf{do} \\
\nl \quad \quad \quad $\mathbf{B}_i,\pmb\Lambda_i,\mathbf{R}_i=\mathbf{J}_{\cdot\cdot i1}, \mathbf{J}_{\cdot\cdot i2}, \mathbf{J}_{\cdot\cdot i3}$ \quad $\left(\mathbf{B}_i,\pmb\Lambda_i,\mathbf{R}_i \in \mathbb{R}^{N\times N}\right)$\\
\nl \quad  \quad \quad $\mathbf{V}_G^{(i)}=\mathbf{B}_i \odot \exp\left[-\pmb\Lambda_i\odot(\mathbf{D}-\mathbf{R}_i) \odot(\mathbf{D}-\mathbf{R}_i)\right]$\\
\nl \quad \ \ \textbf{end for}\\
\lnl{ln: sum of gaussians} \quad \ \ $\textbf{V}_G=\sum_{i=1}^{C_g}\mathbf{V}_{G}^{(i)}$ \\
\lnl{ln: ionic interactions} \quad \ \ $\mathbf{\bar{V}}_G=\mathtt{{Sigmoid}}\left(\mathtt{L}^G(\pmb{z};b_1,\pmb{w}_1)\right)\cdot \textbf{V}_G$ \hfill $b_1\in\mathbb{R}^{1},\pmb{w}_1\in \mathbb{R}^{C_z}$ \\

\nl \quad \ \ \textcolor{olive}{\# Coulomb electrostatic potential as a reciprocal of distance}\\ 
\lnl{ln: Coulomb electrostatic potential} \quad \ \  $\mathbf{V}_C=\mathbf{W}_2 \oslash \mathbf{D}$
% =\mbox{matricize}\left(\mathtt{LinearNoBias}\left(1/\mbox{vec}(\mathbf{D});W_2\right)\right)
\hfill $\mathbf{W}_2\in\mathbb{R}^{N\times N}$\\
\lnl{ln: coulomb} \quad \ \ $\mathbf{\bar{V}}_C=\mathtt{{Sigmoid}}\left(\mathtt{L}^C(\pmb{z};b_2,\pmb{w}_2)\right)\cdot\mathbf{V}_C$ \hfill 
$b_2\in\mathbb{R}^{1},\pmb{w}_2\in \mathbb{R}^{C_z}$ \\
\nl \quad \ \ \textcolor{olive}{\# Total effective potential energy by following eq.~(\ref{eq.V_E})} \\ 
\lnl{ln: sum of potential} \quad \ \ $\mathbf{V}_1 = \mathtt{Dropout}_{0.2}\left(\bar{\mathbf{V}}_G + \bar{\mathbf{V}}_C\right)$ \\
\lnl{ln: feedfoward} \quad \ \ $\mathbf{V}_2 = \mathbf{V}_1 +\mathtt{FeedForward}(\mathbf{V}_1;\pmb{b}_3, \pmb{b}_4,\mathbf{W}_3,\mathbf{W}_4)$ \hfill $\pmb{b}_3\in\mathbb{R}^{hN}, \pmb{b}_4\in\mathbb{R}^{N}$\\
\lnl{ln: potential} \quad \ \ $\Delta \mathbf{G}(\mathbf{S}_0)=\mathtt{BatchNorm}\left(\mathbf{V}_2\right)$ \hfill $\mathbf{W}_3\in\mathbb{R}^{N\times (hN)},\mathbf{W}_4\in\mathbb{R}^{(hN)\times N}$\\
% \nl \quad \ \ \# Contact-specific random effect \\ 
% \nl \quad \ \ $\mathbf{C}_1 = \text{FeedForward}(\mathbf{C};\pmb{b}_3, \pmb{b}_4,\pmb{w}_3,\pmb{w}_4)$ \hfill $\pmb{b}_3\in\mathbb{R}^{hC_t}, \pmb{b}_4\in\mathbb{R}^{C_t},\pmb{w}_3\in\mathbb{R}^{C_t\times (hC_t)},\pmb{w}_4\in\mathbb{R}^{(hC_t)\times C_t}$\\
% \nl \quad \ \ $\mathbf{C}_2 = \text{Dropout}_{0.2}(\mathbf{C}_1)$\\
\nl \quad \ \ \textcolor{olive}{\# Scale parameter model} \\ 
\lnl{ln: sigma} \quad \ \ $\sigma(\pmb{z}) = \mathtt{Softplus}\left(\mathtt{L}^\sigma(\mathbf{\pmb{z}};b_5,\pmb{w}_5)\right)$ \hfill $b_5\in\mathbb{R}^{1}, \pmb{w}_5\in\mathbb{R}^{C_z}$\\
\nl \quad \ \ \textcolor{olive}{\# Lifetime probability model by following eq.~(\ref{eq.H_fun})}  \\ 
\nl \quad \ \ $y = \mathtt{L}(\log t;b_6,w_6)$ \hfill $b_6\in\mathbb{R}^1,w_6\in \mathbb{R}^1$\\
\nl \quad \ \ $\mathbf{H}(t|\mathbf{S}_0; \pmb\theta) = \left[1-\Phi\left(\frac{y-\Delta \mathbf{G}(\mathbf{S}_0)/\mathtt{L}_{\mathbb{T}}(\mathbb{T}; b_6,w_6)}{\sigma(\pmb{z})}\right) \right] \odot \mathbf{M}$ \hfill $b_7\in\mathbb{R}^{1}, w_7\in\mathbb{R}^{1}$\\
\textbf{Return}  $\mathbf{H}(t|\mathbf{S}_0; \pmb\theta)$ \hfill $\pmb\theta$: collection of model parameters \\
\caption{RNA Lifetime Hybrid Network Model (RNA-LifeTime)}\label{algo: model}
\end{algorithm}

% In the context of molecular dynamics simulations, the Boltzmann distribution is used to describe the behavior of a system of atoms. The probability that the molecule is in a state characterized by atomic positions $\pmb{\gamma}$
% \begin{equation}
%     p(\pmb{\gamma})=\frac{1}{Z}\exp\left(-\frac{U(\pmb\gamma)}{k_{B}\mathbb{T}}\right)
% \end{equation}
% where $\mathbb{T}$ is the temperature, $k_B$ is the Boltzmann constant, normalization denominator $Z$ is called canonical partition function and $U(\pmb\gamma)$ is the potential energy. The Boltzmann distribution reveals that temperature plays a crucial role in molecular dynamics (MD) simulations by acting as a threshold. With the rising temperature, the system's potential energy increases, enabling the atomic positions to deviate further from the thermodynamic equilibrium. 
% Take the harmonic bond stretching potential \eqref{eq: harmonic bond potential} for example. The pair distance has probability density function such that $p(r_{ij})\propto \exp\left(-\frac{ \frac{1}{2}k_{ij}(r_{ij}-r_{0,ij})^2}{k_{B}\mathbb{T}}\right)$, At a constant likelihood, the bond length $r_{ij}$ can stretch away from its equilibrium value, $r_{0,ij}$, only when the temperature rises. For the Coulomb potential with the pair distance probability $p(r_{ij})\propto \exp\left(-\frac{q_iq_j}{k_{B}\varepsilon r_{ij}\mathbb{T}}\right)$ implies the distance is reciprocal of temperature.

The free energy barrier for an RNA system refers to the energy difference between the folded native and unfolded states of the sequence \citep{bryngelson1989intermediates,morgan1996evidence}. This barrier is crucial in determining the stability and folding kinetics of the RNA. The magnitude of the free energy barrier depends on various factors, such as the RNA sequence, temperature, and ionic concentration \citep{thirumalai2005rna}. To represent the energy barrier crossing, a sigmoid function is employed to modulate the impact of temperature and ionic concentration on effective potential energies $\mathbf{\bar{V}}_G$ (Algorithm~\ref{algo: model} Lines \ref{ln: ionic interactions}). 

In RNA-LifeTime, we use a mean-field approach to account for the effect of ionic concentrations. In specific, the positively charged ions (Mg\textsuperscript{2+}) are considered to uniformly neutralize the negative charge of all residues, ensuring that each residue carries an averaged negative charge between 0 and 1 (i.e., $0\leq q_1=q_2=\ldots =q_N<1$). 
In the future research, we will consider the unevenly distributed ions.
Consequently, the Coulomb potential described in eq.~\eqref{eq: Coulomb potentials} decreases (but remains greater than or equal to zero) as the concentration of positively charged ions increases, due to the fact that $0\leq q_iq_j <1$. In RNA-LifeTime, a sigmoid function is employed to represent the role of ionic concentration in reducing the Coulomb potential (refer to Algorithm~\ref{algo: model} Line \ref{ln: coulomb}).
Then, based on eq.~(\ref{eq.V_E}), the total effective potential energy of each residue pair is calculated as the sum of Coulomb electrostatic potential and solvent-mediated ionic interactions  (Algorithm~\ref{algo: model} Line \ref{ln: sum of potential}). It follows a two-layer fully-connected feedforward network characterizing the effect between native contacts (Algorithm~\ref{algo: model} Line \ref{ln: feedfoward}).

% In experimental investigations of protein folding thermodynamics, numerous proteins display an all-or-none behavior that resembles a phase transition in a finite system. This observation suggests the presence of a free energy barrier, typically on the order of several $\mathbb{T}$, separating the folded and unfolded states \citep{bryngelson1989intermediates}. To represent the energy barrier, a sigmoid function is employed, which modulates the temperature's impact on potential energy (Algorithm~\ref{algo: model} Lines \ref{ln: ionic interactions} and \ref{ln: coulomb}).
% The model for the scale parameter, denoted as $\sigma(\pmb{z})$, is closely linked to environmental conditions. As the system exhibits greater variability at elevated temperatures and reduced ionic concentrations, the scale model is assumed to be a linear function $\sigma(\pmb{z}) = \text{L}^\sigma(\mathbf{\pmb{z}};b_5,\pmb{w}_5)$ (Algorithm~\ref{algo: model} Line \ref{ln: sigma}).

 The system exhibits increased thermodynamic variability in response to elevated temperatures.
 % and decreased ionic concentrations. 
Therefore, the scale parameter is modeled by a linear function followed by a softplus activation function, i.e., $\sigma(\pmb{z}) = \mathtt{Softplus}\left(\mathtt{L}^\sigma(\mathbf{\pmb{z}};b_5,\pmb{w}_5)\right)$, as shown in Algorithm~\ref{algo: model} Line \ref{ln: sigma}. The softplus function $\mathtt{Softplus}=\log(1+\exp(x))$ ensures the output of the linear function is positive and has a smooth, continuous gradient.

\subsection{Loss Function and Training Procedure}\label{subsec: loss}

We express the lifetime probability model as $\mathbf{H}(t|\mathbf{S}_0; \pmb\theta)$ specified by parameters $\pmb\theta = (b_1,b_2,\ldots,b_7,\mathbf{W}_1,\mathbf{W}_2,\mathbf{W}_3,\mathbf{W}_4,\pmb{w}_1,\pmb{w}_2,\pmb{w}_5,w_6,w_7)$. Given the trajectory observations, denoted by
$\mathcal{D}=\{\left(\mathbf{S}_{0,n},\mathbf{Y}_n(t_\ell)\right)| \ell\in[L],n\in[N_s]\}$, where the output trajectory $\mathbf{Y}(t)$ is measured at discrete time $t_\ell$ for $\ell\in[L]$. Here $N_s$ is the total number of simulation trajectories, and $L$ is the number of timestep records in each trajectory.
The 
 model is trained by minimizing mean absolute error (MAE, $p=1$) with each scenario representation, %the per-sample loss defined as follows 
 \begin{equation}\label{eq: loss function}
     \mathrm{Loss}_p\left(\mathbf{H}(t_\ell|\mathbf{S}_0; \pmb\theta),\mathbf{Y}(t_\ell)\right)=\frac{1}{|\mathbb{C}(\mathbf{X}_0)|}\sum_{(i,j)\in \mathbb{C}(\mathbf{X}_0)}\Vert H_{ij}(t_\ell|\mathbf{S}_0;\pmb\theta)-\mathbf{Y}_{ij}(t_\ell)\Vert_p.
 \end{equation}
 % where $B$ represents the mini-batch size for each training optimization step.

 In the initial phases of development, we found that degradation displayed an all-or-none behavior. That means once triggered, the degradation process quickly finishes: the fraction of native contacts stops decreasing within the first 25 timesteps before fluctuating around a constant. This observation implies that samples from earlier times contain more valuable information about the degradation process (referred to as "positive samples"). Consequently, we implemented a technique to upsample data collected in the early stages of the process (i.e., when $\ell \leq U$), where $U$ denotes the upsampling threshold. We define the sampling probability as,
\begin{equation}\label{eq: upsampling}
    P(\ell) = \begin{cases}
    (U - \ell+2) / \left(\frac{U(U+1)}{2} + L\right) & \text{if $\ell \leq U$}\\
    1 / \left(\frac{U(U+1)}{2} + L\right) & \text{if $\ell > U$}
\end{cases}
\end{equation}
which assigns a higher weight to samples closer to the starting time. It is also easy to verify that $\sum_{\ell=1}^L P(\ell)=1$. This approach substantially enhanced prediction accuracy.

% The loss function was hand-selected and underwent minimal fine-tuning during the development process. There is potential for enhanced accuracy through more comprehensive optimization of the loss function selection and assigning slightly higher weights to early time steps ($\ell \leq 25$). 

\begin{algorithm}[hbt!]%[H]
\SetAlgoLined
\LinesNumbered
\textbf{Input}: Dataset $\mathcal{D}=\{\left(\mathbf{S}_{0,n},\mathbf{Y}_n(t_\ell)\right)| \ell\in[L],n\in[N_s]\}$
where the $n$-th trajectory input $\mathbf{S}_{0,n}$ includes the pair distance matrix $\mathbf{D}_n\in\mathbb{R}^{N\times N}$, native contact tensor $\mathbf{C}_n\in\mathbb{R}^{N\times N\times C_g}$, $\mathbf{M}_n\in \{0,1\}^{N\times N}$, environmental conditions $\pmb{z}_n\in\mathbb{R}^{C_z}$, 
upsample threshold $U$, the number of epochs $K$, and mini-batch size $B$. %upsample threshold $U=25$, the number of epochs $K=15$ and mini-batch size $B=512$.
\\
% \textbf{Output}: $\mathbf{H}(t|\mathbf{S}_0)\in [0,1]^{N\times N}$\\
\nl \textbf{for all} $k \in \{1,2,\ldots,K\}$ \textbf{do}\\
\nl \quad\quad \textbf{for all} $h \in \{1,2,\ldots,\lceil LN_s / B\rceil\}$ \textbf{do}\\
\nl \quad\quad\quad\quad Randomly sample a mini-batch $\mathcal{B}=\{(n,\ell)_b\}_{b=1}^B$ from the dataset $\mathcal{D}$ by applying \eqref{eq: upsampling} \\
\nl \quad\quad  \quad\quad \textbf{for all} $(n,\ell) \in \{(n,\ell)_b\}_{b=1}^B$ \textbf{do}\\
\nl \quad\quad \quad\quad \quad\quad$\mathbf{H}_n(t_\ell|\mathbf{S}_0; \pmb\theta)=$LifetimeModel($t_\ell$, $\mathbf{D}_n$, $\mathbf{C}_n$, $\mathbf{M}_n$, $\pmb{z}_n$)\\
\nl \quad\quad \quad\quad \textbf{end for}\\
\nl \quad\quad \quad\quad  $\mathcal{L}(\pmb\theta)=\frac{1}{|\mathcal{B}|}\sum_{(n,\ell)\in \mathcal{B}}\mathrm{Loss}_1\left(\mathbf{H}_n(t_\ell|\mathbf{S}_0; \pmb\theta),\mathbf{Y}_n(t_\ell)\right)$ by applying (\ref{eq: loss function}) \\
\nl \quad\quad \quad\quad  $\pmb\theta_{k}\leftarrow \text{Adam}(\mathcal{L}(\pmb\theta_{k-1}))$\\
\textbf{Return}  $\pmb\theta_K$
\caption{Model Training Procedure}\label{algo: training}
\end{algorithm}

The min-batch gradient descent approach is used to train the RNA-LifeTime hybrid model; see the procedure in Algorithm~\ref{algo: training}. 
This algorithm iteratively trains the model over multiple epochs, where each epoch generates mini-batches of the dataset by applying the upsampling probability in eq.~\eqref{eq: upsampling}. Then, a sample average approximate (SAA) of the expected loss function is computed to compare the predictions to the actual lifetime probabilities. In the training, we use Adam optimizer
with a base learning rate $0.003$ and coefficients used for computing running averages of gradient and its square (default: (0.9, 0.999)) %$\beta_1 = 0.9$, $\beta_2=0.999$ 
to search for the best fit parameters $\pmb\theta$ for the RNA-LifeTime model. 
By default, the weights of the Linear layers are initialized using the LeCun (fan-in) initialization strategy \citep{orr1998neural}.
\section{Empirical Study}\label{sec: empirical study}

In this section, we study the performance of RNA-LifeTime of  multiple single RNA molecules in their degradation processes.  We use the classic AFT model with the linear mean function (AFT-Linear) as the baseline. In specific, the input vector for AFT-Linear includes the feature mapping that concatenates all vectorized inputs $\pmb{s}_0$, and the scale parameter model $\sigma(\pmb{z})$ is the same as RNA-LifeTime. 
% Although structurally simple, this model contains 295,627,971 trainable parameters in total (more than 7000 times larger than RNA-LifeTime).

\subsection{Data Sources and Training Procedure}

The RNA molecules (14 in total) are manually selected from PDB database \citep{wwpdb2019protein}, and preprocessed using SMOG 2 \citep{noel2016smog}. The preprocessing includes the steps: (1) remove water molecules and hydrogen atoms (only consider heavy atoms); (2) add  Mg\textsuperscript{2+}  and K\textsuperscript{+} ions 
%to the simulation system; (3) 
and calculate the total charge of the molecular simulation system; and (3) neutralize the system with Cl\textsuperscript{-} ions. 
% The native contacts were determined by SMOG using 

The MD simulations were generated by utilizing OpenSMOG \citep{de2022smog} with the force field calculated by ``AA\_ions\_Wang22.v1" \citep{wang2022diffuse}. Simulation experiments were conducted under eight reduced temperatures (r.t.) ranging from 0.5 to 1.2 and three Mg\textsuperscript{2+} concentrations (0.1mM, 1mM and 10mM). Here, 0.5 r.t. corresponds to room temperature, and 1.0 r.t. corresponds to 350K. For each combination of RNA conformation, temperature, and Mg\textsuperscript{2+} concentration, we conducted 10 simulation replications, each taking 0.5 hours per GPU. 
The MD simulations were conducted using 6-8 GPUs and required approximately one week to complete, i.e., $8 \times 3 \times 14 \times 10 \times 0.5 = 1680$ hours/GPU, where 8 is the number of temperature levels, 3 is the number of Mg\textsuperscript{2+} concentration levels,  and 14 is the number of RNA molecules.
After MD simulations, the simulated trajectories were post-processed to generate pair distance matrices, contact type, environmental features, and the fractions of native contacts. Then they were utilized to train and evaluate the performance of proposed RNA-LifeTime hybrid model. %then input to the RNA-LifeTime.

MD simulation trajectories were padded to 256 residues. 
% For each trajectory, we selected every other timestep within the first 1000 timesteps (i.e., $L=500$ and $t_L=1000$) to generate the training outputs.
For each trajectory, we used the initial state
of the RNA system $\mathbf{S}_0$ as the input and selected the lifetime probability 
% (eq.~\eqref{eq: probability of lifetime of native contacts})
% fraction of native contacts 
$\mbox{Pr}(T_{ij}>t|\mathbf{X}_0,\pmb{z})$ for each $(i,j)\in\mathbb{C}(\mathbf{X}_0)$ at every other timestep within the first 1000 timesteps as training outputs, thus utilizing 500 timesteps in total (i.e., $L=500$).
Then we selected five sequences %(PDB IDs: 1ATO, 2LPS, 2KPV, 1RFR, and 1I4C)
at random and designated their trajectories as the test set. The remaining nine sequences were assigned to the training and validation sets. We trained the model on a cluster of 64 CPUs and 128 GB memory until convergence (i.e., the validation loss stops decreasing after 5 epochs). 
The mini-batch size, epochs, and upsample threshold were selected to be $B = 512$, $K=20$, and $U = 25$ (Algorithm~\ref{algo: training}). 

\subsection{Model Performance}
The expected lifetime $T_{ij}>0$ can be calculated as follows: $\E[T_{ij}|\mathbf{S}_0]=\int_0^\infty \mbox{Pr}(T_{ij} > t|\mathbf{S}_0) \mbox{d} t$. 
The expected RNA lifetime from MD simulations can be approximated by $\E^{MD}[T_{ij}|\mathbf{S}_0]=\int_0^\infty Q(t|\mathbf{S}_0) \mbox{d} t\approx\sum_{\ell}^L Q(t_\ell|\mathbf{S}_0)\Delta t$ with $\Delta t = t_\ell - t_{\ell-1}$. Similarly, Similarly, the expected RNA lifetime predicted by RNA-LifeTime becomes $\E^{RNA-LT}[T_{ij}|\mathbf{S}_0]=\int_0^\infty H_{ij}(t|\mathbf{S}_0) \mbox{d} t\approx\sum_{\ell}^L H_{ij}(t_\ell|\mathbf{S}_0)\Delta t$. Then we define the mean absolute error of expected lifetime (MAE-LT) between MD simulation and RNA-LifeTime prediction as
\begin{equation*}
    \mbox{MAE-LT}=\frac{1}{N_s}\sum_{n=1}^{N_s}\frac{1}{|\mathbb{C}(\mathbf{X}_{0,n})|}\sum_{(i,j)\in\mathbb{C}(\mathbf{X}_{0,n})}\left|\E^{MD}[T_{ij}|\mathbf{S}_0] - \E^{RNA-LT}[T_{ij}|\mathbf{S}_0]\right|,
\end{equation*}
where $N_s$ is the total number of simulation trajectories. Overall, the performances are evaluated by three metrics:  MAE-LT, MAE (eq.~\eqref{eq: loss function} with $p=1$), and mean squared error (MSE) (eq.~\eqref{eq: loss function} with $p=2$).

\begin{figure}[bht!]
	\centering
\includegraphics[width=1\textwidth]{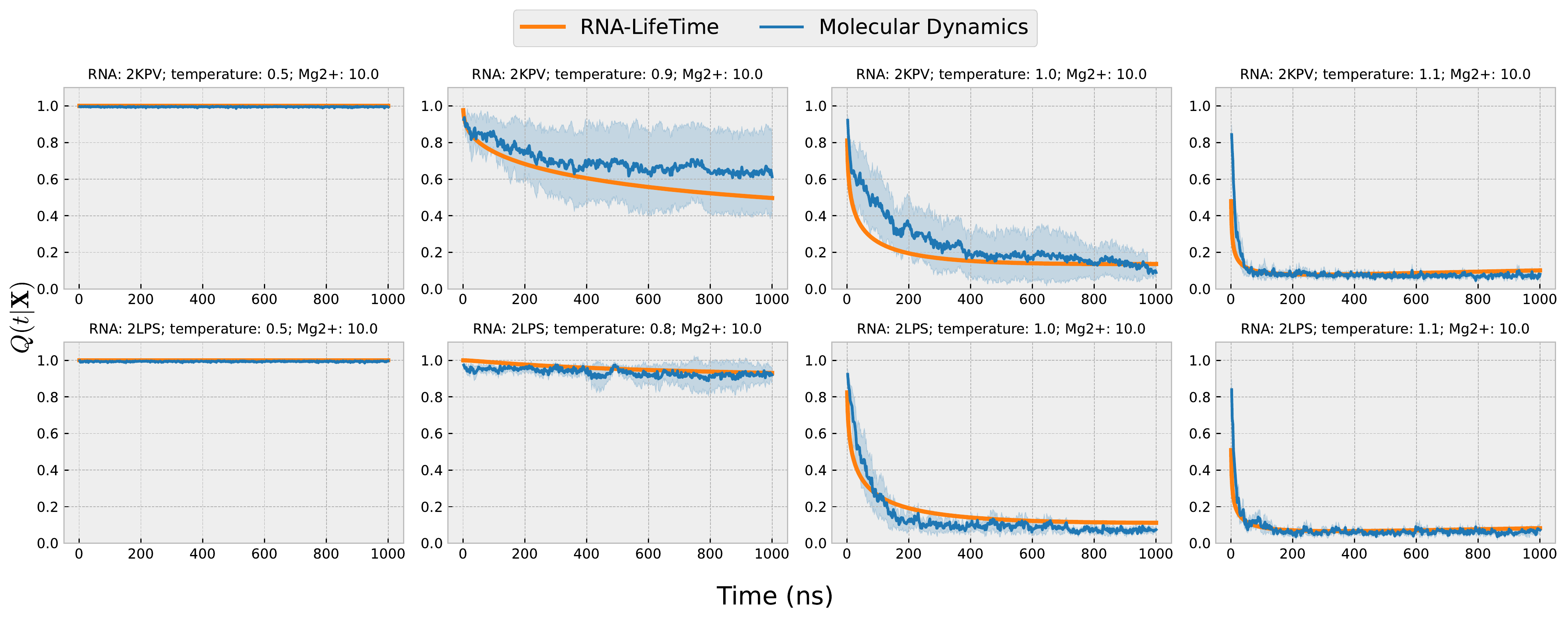}
	\caption{Predicted (orange) vs. MD simulated (blue) trajectories for a Let-7  miRNA  (PDB 1ATO, top row) and Yeast ai5(gamma) Group II Intron (PDB 2LPS, bottom row). Error bars indicate the confidence intervals for the fractions of native contacts, as derived from 10 replicate simulations.} 
	\label{fig: prediction result}
\end{figure}

RNA-LifeTime accurately predicts native contact fractions for RNA molecules, as illustrated in Figure~\ref{fig: prediction result}. Since it was trained on a small set of MD simulation data containing only nine RNA molecules, whereas AlphaFold 2 was trained on a large dataset of 170,000 protein structures, we anticipate that the accuracy could be improved by including more simulations of additional RNA molecules in the training set.

\begin{table}[ht]
\caption{RNA-LifeTime prediction performance on RNA degradation rate. Test errors are summarized as the mean (standard deviation) across 3 macro-replications.
}\label{table: performance}
\centering
\scalebox{0.95}{
\begin{tabular}{@{}l|ccccc@{}}
\toprule
             & \# Parameters & MAE-LT       & MAE    & MSE  & Training Time \\ \midrule
AFT-Linear   & 295,627,971   &     262.7 (10.2)         & 0.3744 (0.013) &   0.300 (0.012)  & $\approx$ 1 day \\
RNA-LifeTime ($C_g=1$)& 42,246        & 67.84 (2.8) & 0.1531 (0.003) &
0.106 (0.003) & $\approx$ 2 Hours \\
RNA-LifeTime ($C_g=3$)& 42,306       & 70.24 (2.2) & 0.1461 (0.002) &
0.093 (0.002) & $\approx$ 3 Hours \\
RNA-LifeTime ($C_g=5$)& 42,366        & 67.44 (2.1) & 0.1461 (0.002) &
0.090 (0.002) & $\approx$ 4 Hours \\\midrule
MD-Average & - & 52.5 & 0.1238 & 0.067& $\approx$ 1 weeks (using 6-8 GPUs)\\ \bottomrule
\end{tabular}}
\end{table}

The results in Table~\ref{table: performance} demonstrate that the proposed RNA-LifeTime hybrid model can achieve high accuracy in the lifetime probability prediction.
Here the MD simulation is supposed as the ground truth model.  RNA-LifeTime performs better than AFT-Linear in terms of MAE-LT, MAE, and MSE, while having much fewer parameters and requiring a shorter training time. This suggests that RNA-LifeTime is more efficient and accurate for predicting RNA degradation rate. Among the RNA-LifeTime models tested, the one with $C_g=5$ achieves the best performance, with an MAE-LT of 67.44 (2.1), a MAE of 0.1461 (0.002), and an MSE of 0.090 (0.002).
%in approximately 4 hours of training time. 

Furthermore, we compute the MAE-LT, MAE, and MSE for the mean trajectories of the ground truth MD simulation (MD-average) to represent the intrinsic randomness in the system (refer to the last row of Table~\ref{table: performance}). These results approximate the lowest achievable error. In comparison, the estimation errors of RNA-LifeTime are close to those of the MD-average, suggesting that RNA-LifeTime demonstrates a promising performance.

\section{Conclusion}\label{sec: conclusion}

The proposed hybrid model for RNA structure and functional dynamics builds on the scientific understanding of biomolecular interactions and it allows us to efficiently predict molecular conformational changes while providing insights into the energetics and dynamics of an enzymatic reaction network. Our RNA-LifeTime approach is capable of predicting RNA lifetime with high accuracy in  short time, unlike traditional methods that may require a significantly longer time of molecular dynamics  simulations. This modeling strategy can be extended to enzymatic molecular reaction networks and facilitate the development of multi-scale bioprocess digital twin. Further coupling with process analytical technologies and optimal design of experiments, we can accelerate new drug discovery and manufacturing process development.

\bibliographystyle{unsrtnat}
\bibliography{references} 

\begin{thebibliography}{19}
\providecommand{\natexlab}[1]{#1}
\providecommand{\url}[1]{\texttt{#1}}
\expandafter\ifx\csname urlstyle\endcsname\relax
  \providecommand{\doi}[1]{doi: #1}\else
  \providecommand{\doi}{doi: \begingroup \urlstyle{rm}\Url}\fi

\bibitem[Jumper et~al.(2021)Jumper, Evans, Pritzel, Green, Figurnov,
  Ronneberger, Tunyasuvunakool, Bates, {\v{Z}}{\'\i}dek, Potapenko,
  et~al.]{jumper2021highly}
John Jumper, Richard Evans, Alexander Pritzel, Tim Green, Michael Figurnov,
  Olaf Ronneberger, Kathryn Tunyasuvunakool, Russ Bates, Augustin
  {\v{Z}}{\'\i}dek, Anna Potapenko, et~al.
\newblock Highly accurate protein structure prediction with alphafold.
\newblock \emph{Nature}, 596\penalty0 (7873):\penalty0 583--589, 2021.

\bibitem[Xie and Pedrielli(2022)]{xie2022discovery}
Wei Xie and Giulia Pedrielli.
\newblock From discovery to production: Challenges and novel methodologies for
  next generation biomanufacturing.
\newblock In \emph{2022 Winter Simulation Conference (WSC)}, pages 238--252.
  IEEE, 2022.

\bibitem[\v{S}poner et~al.(2018)\v{S}poner, Bussi, Krepl, Baná\v{s}, Bottaro,
  Cunha, Gil-Ley, Pinamonti, Poblete, Jure\v{c}ka, et~al.]{sponer2018rna}
Ji\v{r}{\'\i} \v{S}poner, Giovanni Bussi, Miroslav Krepl, Pavel Baná\v{s},
  Sandro Bottaro, Richard~A Cunha, Alejandro Gil-Ley, Giovanni Pinamonti,
  Sim{\'o}n Poblete, Petr Jure\v{c}ka, et~al.
\newblock {RNA} structural dynamics as captured by molecular simulations: A
  comprehensive overview.
\newblock \emph{Chemical reviews}, 118\penalty0 (8):\penalty0 4177--4338, 2018.

\bibitem[Wang et~al.(2022)Wang, Levi, Mohanty, and Whitford]{wang2022diffuse}
Ailun Wang, Mariana Levi, Udayan Mohanty, and Paul~C Whitford.
\newblock Diffuse ions coordinate dynamics in a ribonucleoprotein assembly.
\newblock \emph{Journal of the American Chemical Society}, 144\penalty0
  (21):\penalty0 9510--9522, 2022.

\bibitem[Whitford et~al.(2010)Whitford, Onuchic, and
  Sanbonmatsu]{whitford2010connecting}
Paul~C Whitford, Jos{\'e}~N Onuchic, and Karissa~Y Sanbonmatsu.
\newblock Connecting energy landscapes with experimental rates for
  aminoacyl-trna accommodation in the ribosome.
\newblock \emph{Journal of the American Chemical Society}, 132\penalty0
  (38):\penalty0 13170--13171, 2010.

\bibitem[Vill{\`a} and Warshel(2001)]{Jordi_2001}
Jordi Vill{\`a} and Arieh Warshel.
\newblock Energetics and dynamics of enzymatic reactions.
\newblock \emph{The Journal of Physical Chemistry B}, 105\penalty0
  (33):\penalty0 7887--7907, Aug 2001.
\newblock ISSN 1520-6106.
\newblock \doi{10.1021/jp011048h}.
\newblock URL \url{https://doi.org/10.1021/jp011048h}.

\bibitem[Noel et~al.(2012)Noel, Whitford, and Onuchic]{noel2012shadow}
Jeffrey~K Noel, Paul~C Whitford, and Jos{\'e}~N Onuchic.
\newblock The shadow map: a general contact definition for capturing the
  dynamics of biomolecular folding and function.
\newblock \emph{The journal of physical chemistry B}, 116\penalty0
  (29):\penalty0 8692--8702, 2012.

\bibitem[Wang et~al.(2019)Wang, Williams, Chirasani, Krokhotin, Das, and
  Dokholyan]{wang2019limits}
Jian Wang, Benfeard Williams, Venkata~R Chirasani, Andrey Krokhotin, Rajeshree
  Das, and Nikolay~V Dokholyan.
\newblock Limits in accuracy and a strategy of rna structure prediction using
  experimental information.
\newblock \emph{Nucleic acids research}, 47\penalty0 (11):\penalty0 5563--5572,
  2019.

\bibitem[Srivastava et~al.(2014)Srivastava, Hinton, Krizhevsky, Sutskever, and
  Salakhutdinov]{srivastava2014dropout}
Nitish Srivastava, Geoffrey Hinton, Alex Krizhevsky, Ilya Sutskever, and Ruslan
  Salakhutdinov.
\newblock Dropout: a simple way to prevent neural networks from overfitting.
\newblock \emph{The journal of machine learning research}, 15\penalty0
  (1):\penalty0 1929--1958, 2014.

\bibitem[Ioffe and Szegedy(2015)]{ioffe2015batch}
Sergey Ioffe and Christian Szegedy.
\newblock Batch normalization: Accelerating deep network training by reducing
  internal covariate shift.
\newblock In \emph{International conference on machine learning}, pages
  448--456. pmlr, 2015.

\bibitem[Best et~al.(2013)Best, Hummer, and Eaton]{best2013native}
Robert~B Best, Gerhard Hummer, and William~A Eaton.
\newblock Native contacts determine protein folding mechanisms in atomistic
  simulations.
\newblock \emph{Proceedings of the National Academy of Sciences}, 110\penalty0
  (44):\penalty0 17874--17879, 2013.

\bibitem[Lee and Wang(2003)]{lee2003statistical}
Elisa~T Lee and John Wang.
\newblock \emph{Statistical methods for survival data analysis}, volume 476.
\newblock John Wiley \& Sons, 2003.

\bibitem[Bryngelson and Wolynes(1989)]{bryngelson1989intermediates}
Joseph~D Bryngelson and Peter~G Wolynes.
\newblock Intermediates and barrier crossing in a random energy model (with
  applications to protein folding).
\newblock \emph{The Journal of Physical Chemistry}, 93\penalty0 (19):\penalty0
  6902--6915, 1989.

\bibitem[Morgan and Higgs(1996)]{morgan1996evidence}
Steven~R Morgan and Paul~G Higgs.
\newblock Evidence for kinetic effects in the folding of large rna molecules.
\newblock \emph{The Journal of chemical physics}, 105\penalty0 (16):\penalty0
  7152--7157, 1996.

\bibitem[Thirumalai and Hyeon(2005)]{thirumalai2005rna}
D~Thirumalai and Changbong Hyeon.
\newblock Rna and protein folding: common themes and variations.
\newblock \emph{Biochemistry}, 44\penalty0 (13):\penalty0 4957--4970, 2005.

\bibitem[Orr and M{\"u}ller(1998)]{orr1998neural}
Genevieve~B Orr and Klaus-Robert M{\"u}ller.
\newblock \emph{Neural networks: tricks of the trade}.
\newblock Springer, 1998.

\bibitem[wwPDB consortium(2018)]{wwpdb2019protein}
wwPDB consortium.
\newblock {Protein Data Bank: the single global archive for 3D macromolecular
  structure data}.
\newblock \emph{Nucleic Acids Research}, 47\penalty0 (D1):\penalty0 D520--D528,
  10 2018.
\newblock ISSN 0305-1048.
\newblock \doi{10.1093/nar/gky949}.
\newblock URL \url{https://doi.org/10.1093/nar/gky949}.

\bibitem[Noel et~al.(2016)Noel, Levi, Raghunathan, Lammert, Hayes, Onuchic, and
  Whitford]{noel2016smog}
Jeffrey~K. Noel, Mariana Levi, Mohit Raghunathan, Heiko Lammert, Ryan~L. Hayes,
  José~N. Onuchic, and Paul~C. Whitford.
\newblock Smog 2: A versatile software package for generating structure-based
  models.
\newblock \emph{PLOS Computational Biology}, 12\penalty0 (3):\penalty0 1--14,
  03 2016.
\newblock \doi{10.1371/journal.pcbi.1004794}.
\newblock URL \url{https://doi.org/10.1371/journal.pcbi.1004794}.

\bibitem[de~Oliveira~Jr et~al.(2022)de~Oliveira~Jr, Contessoto, Hassan, Byju,
  Wang, Wang, Dodero-Rojas, Mohanty, Noel, Onuchic, and Whitford]{de2022smog}
Antonio~B. de~Oliveira~Jr, Vinícius~G. Contessoto, Asem Hassan, Sandra Byju,
  Ailun Wang, Yang Wang, Esteban Dodero-Rojas, Udayan Mohanty, Jeffrey~K. Noel,
  Jose~N. Onuchic, and Paul~C. Whitford.
\newblock Smog 2 and opensmog: Extending the limits of structure-based models.
\newblock \emph{Protein Science}, 31\penalty0 (1):\penalty0 158--172, 2022.
\newblock \doi{https://doi.org/10.1002/pro.4209}.
\newblock URL \url{https://onlinelibrary.wiley.com/doi/abs/10.1002/pro.4209}.

\end{thebibliography}

\end{document}